\documentclass[aps,floatfix,preprint,showpacs,preprintnumbers,nofootinbib,superscriptaddress,natbib]{revtex4}

\usepackage{graphicx,float}
\usepackage{epsfig}		
\usepackage{dcolumn}
\usepackage{bm}
\usepackage{subfigure}

\usepackage[all]{xy}
\usepackage{amsmath,upgreek}
\usepackage{amssymb}

\usepackage{pdfpages}
\usepackage{color}
\usepackage{graphicx,epstopdf}
\usepackage[colorlinks,hyperindex]{hyperref}

\def\0{\mbox{\tiny $0$}}
\def\1{\mbox{\tiny $1$}}
\def\2{\mbox{\tiny $2$}}
\def\3{\mbox{\tiny $3$}}
\def\4{\mbox{\tiny $4$}}
\def\5{\mbox{\tiny $5$}}
\def\6{\mbox{\tiny $6$}}
\def\7{\mbox{\tiny $7$}}
\def\8{\mbox{\tiny $8$}}
\def\9{\mbox{\tiny $9$}}

\def\f14{\mbox{\tiny $\frac{1}{4}$}}

\def\L{\mbox{\tiny $L$}}

\def\R{\mbox{\tiny $R$}}

\def\mi{\mbox{\tiny $-$}}

\def\mt#1{\mbox{\textsl{#1}}}
\def\bm#1{\mbox{\boldmath$#1$}}

\def\bb#1{\mbox{\footnotesize $(#1)$}}

\begin{document}

\title{Lorentz Invariant Quantum Concurrence for $SU(2) \otimes SU(2)$ spin-parity states}
\author{Alex E. Bernardini}
\email{alexeb@ufscar.br}
\affiliation{~Departamento de F\'{\i}sica, Universidade Federal de S\~ao Carlos, PO Box 676, 13565-905, S\~ao Carlos, SP, Brasil.}
\author{Victor A. S. V. Bittencourt}
\email{victor.bittencourt@mpl.mpg.de}
\affiliation{~Max Planck Institute for the Science of Light, Staudtstra\ss e 2, PLZ 91058, Erlangen, Germany.}
\author{Massimo Blasone}
\email{blasone@sa.infn.it}
\affiliation{Dipartimento di Fisica, Universit\`a degli Studi di Salerno, Via Giovanni Paolo II, 132 84084 Fisciano, Italy.}
\affiliation{Also at: INFN Sezione di Napoli, Gruppo collegato di Salerno, Italy.}

\date{\today}
\renewcommand{\baselinestretch}{1.4}

\begin{abstract}
The quantum concurrence of $SU(2) \otimes SU(2)$ spin-parity states is shown to be invariant under $SO(1,3)$ Lorentz boosts and $O(3)$ rotations when the density matrices are constructed in consonance with the covariant probabilistic distribution of Dirac massive particles.
Similar invariance properties are obtained for the quantum purity and for the trace of unipotent density matrix operators. 
The reported invariance features -- obtained in the scope of the $SU(2) \otimes SU(2)$ corresponding to just one of the inequivalent representations enclosed by the $SL(2,\mathbb{C})\otimes SL(2,\mathbb{C})$ symmetry -- set a more universal and kinematical-independent meaning for the quantum entanglement encoded in systems containing not only information about spin polarization but also the correlated information about intrinsic parity. 
Such a covariant framework is used for computing the Lorentz invariant spin-parity entanglement of spinorial particles coupled to a magnetic field, through which the extensions to more general Poincar\'e classes of spinor interactions are straightforwardly depicted.
\end{abstract}

\date{\today}
\maketitle

\section{Introduction}

The interpretation of quantum information paradigms under the light of relativistic quantum mechanics has been predominantly concerned with how spin-spin and spin-momentum entanglement does change under Lorentz boosts, in particular, in the context of describing communication schemes in the relativistic framework \cite{relat01, relat02, relat03, relat04, relat05, relat06, relat07, relat08, relatvedral}.
It involves, for instance, clock synchronization systematics \cite{clock}, position verification algorithms \cite{position}, teleportation protocols \cite{teleport}, and theoretical issues related to the entanglement encoded by a pair of spin one-half particles \cite{PRA2108}.
In such a relativistic quantum information framework, the fundamental underlying question is to address the influence of relativistic frame transformations onto quantum entanglement to obtain a factorization of kinematical effects from those related to discrete variable quantum correlations, such as spin-spin or spin-parity entanglement \cite{SU2, extfields, diraclike01, diraclike02}.

For quantum correlated states described by the fundamental representation of the $SU(2) \otimes SU(2)$ symmetry, in particular, where $SU(2)$ is constructed as a subset of $SL(2, \mathbb{C})$, which is homomorphic to the homogeneous Lorentz group $SO(1,3)$ (see Appendix), the above issues can be indeed formulated according to the Lorentz covariant framework.
This structure reflects the $SU(2) \otimes SU(2)$ spin-parity intrinsic properties exhibited, for instance, by the free particle solutions of the Dirac equation, which has an associated Hamiltonian\footnote{The free particle Dirac Hamiltonian in the coordinate space reads
$$\label{diracequation}
{H} \, \psi\bb{t,\bm{x}} = i \frac{\partial \, \psi\bb{t,\bm{x}}}{\partial t} = (-i\bm{\nabla} \cdot {\bm{\alpha}} + m {\beta}) \,\psi\bb{t,\bm{x}} = (-i {\alpha}_i\partial^i + m {\beta}) \,\psi\bb{t,\bm{x}} = \pm E_p \,\psi\bb{t,\bm{x}},
$$
where ${H}$ expressed in natural units, i.e. with $c = \hbar = 1$, the eigenvalues, $\pm E_p$, are expressed by $E_p = \sqrt {p^2 + m^2}$, and the matrix operators ${\bm{\alpha}} =({\alpha}_x,\,{\alpha}_y,\,{\alpha}_z)$ and ${\beta}$ satisfy the anticommuting relations, $
{\alpha}_i {\alpha}_j + {\alpha}_j {\alpha}_i = 2 \delta_{ij} {I}_4$, and ${\alpha}_i {\beta} + {\beta} {\alpha}_i =0$, for $i,j = x,y,z$, with $
{\beta}^2 = {I}_4$, where $I_N$ the $N$-dim identity matrix.
These so-called Dirac matrices, ${\alpha}_i$ and ${\beta}$, in their standard (or Dirac) representation are decomposed into tensor products of Pauli matrices \cite{SU2}, as ${\alpha}_i = {\sigma}_x^{(P)} \otimes {\sigma}_i^{(S)}$, for $i = x,y,z$ and ${\beta} = {\sigma}_z^{(P)} \otimes {I}_{2}^{(S)}$.} written in terms of Kronecker products between Pauli matrices, ${\sigma}_i$, as
\begin{equation}
\label{twoqubithamiltonian}
{H} = \bm{p}\cdot ({\sigma}_x^{(P)} \otimes {\bm{\sigma}}^{(S)}) + m ( {\sigma}_z ^{(P)} \otimes {I}_{2}^{(S)}),
\end{equation}
where ${\bm{\sigma}} =({\sigma}_x,\,{\sigma}_y,\,{\sigma}_z)$, and the superscripts $S$ and $P$ refer to the {\em qubits} of spin and parity, and from which, according to the information content interpretation of quantum mechanics, the ${H}$ eigenstates can be identified as two {\em qubit} states whose dynamics is constrained by continuous variables.
This picture supports the construction of Dirac state vectors as double-doublet representations of the $SU(2)\otimes SU(2)$ group,
$ \psi^{\dagger}  =\left(
\psi _{\L\1},\,
\psi _{\L\2},\,
\psi _{\R\1},\,
\psi _{\R\2}\right)
\equiv (\bm{2},\bm{1})\oplus(\bm{1},\bm{2})$,
which is not unique.
In particular, one has the {\em left-handed} {\em spinors} as doublets ($2$-dim representations) of the $SU(2)$ (left) and as singlets ($1$-dim representations) of the $SU(2)$ (right),
$(\bm{2},\bm{1}) \equiv \psi^{\dagger} _{\L} =\left(
\psi _{\L\1},\,
\psi _{\L\2}\right)$ and, of course, the {\em right-handed} {\em spinors} as doublets of the $SU(2)$ (right) and as singlets of the $SU(2)$ (left), $(\bm{1},\bm{2}) \equiv \psi^{\dagger} _{\R} =\left(
\psi _{\R\1},\,
\psi _{\R\2}\right)$.

Within this framework, the normalized stationary eigenstates in the momentum coordinate are written in terms of a sum of direct products which, in the case of the Hamiltonian Eq.~(\ref{twoqubithamiltonian}), and even for extensions involving interacting external fields \cite{extfields}, describes {\em spin-parity} entangled states.
The point here is the quantification of this entanglement properties in a completely covariant way which, as it shall be demonstrated, can be achieved through the identification of the invariance of quantum concurrence under homogeneous Lorentz transformations.

This manuscript is thus devised as follows. Sec. II is devoted to the understanding of the covariant quantum information structures involving Dirac spinor states. The definition of density matrix operators which preserve a trace unipotent invariant behavior under Lorentz boosts is introduced in order to provide the elementary tools for discussing intrinsic quantum correlations in a framework which is driven by the quantum mechanics paradigms.
Sec. III summarizes the main result of this manuscript: from the setup definitions introduced by the covariant framework from Sec. II, a Lorentz invariant expression for the quantum concurrence between spin and intrinsic parity is straightforwardly obtained and applied for discussing spin-parity quantum entanglement for Dirac spinor states coupled to a magnetic field. The quantitative analysis is performed for pure Hamiltonian eigenstates and rank 2 statistical mixtures corresponding to helicity and parity projected states.
Our conclusions are drawn in Sec. IV.

\section{Spinors and Covariant Quantum Information Structures} 

Let one starts from the $3$-dim Euclidean space definition of two component spinors 
\begin{equation}
\chi_{\1}\equiv \chi_{\1} \bb{\hat{\bm{k}}}= \left(\begin{array}{r} \cos(\vartheta/2) \\ e^{+i\varphi}\sin(\vartheta/2) \end{array}
\right) \qquad\mbox{e} \qquad
\chi_{\2}\equiv \chi_{\2} \bb{\hat{\bm{k}}} = \left(\begin{array}{r} -e^{-i\varphi}\sin(\vartheta/2)\\ \cos(\vartheta/2)  \end{array}\right), 
\label{4035B}
\end{equation}
where $\vartheta$ and $\varphi$ correspond to polar and azimutal angles, respectively, and the spherical unitary vector  $\hat{\bm{k}}$ is denoted by $\hat{\bm{k}} = \cos(\varphi) \sin(\vartheta)\hat{x} + \sin(\varphi)\sin(\vartheta)\hat{y} + \cos(\vartheta)\hat{z}$.

Spinors as well as geometrical vectors and more general tensorial objects have a characteristic group representation which is defined by the way that they transform with respect to the algebra of some symmetry group.
Invariant quantities are obtained through elementary operations involving elements of these group representations according to the corresponding algebra properties.
For instance, under $O(3)$ group transformations, which describe the $3$-dim Euclidean rotations reflected by a gradual and continuous modification of the spherical angles, from $(\vartheta_{\1},\varphi_{\1})$ to $(\vartheta_{\2},\varphi_{\2})$ onto the expression for $\chi_{\1,\2}$, one notices that the unchanged expression for $\chi^{\dagger}_{s} \chi_{r} = \delta_{sr}$ is an $O(3)$ invariant quantity.

For a more consistent understanding of the above spinor properties, in what concern their enhancement with space-time properties, a similar notion of invariance must be extended to the $(1+3)$-dim Minkowski space.
It is indeed performed by generalizing the meaning of the symmetries involved into $O(3)$ rotations to homogeneous Lorentz transformations of the $SO(1,3)$ which have a role analogous to the $O(3)$ rotations in the discussion of the {\em spinor} properties.
This is performed by the identification of $\chi_{\1,\2}$ as the fundamental $2$-dim representations, $\bf{2}$, of the algebra of the $SU(2)$, which is related to the $SO(1,3)$ symmetry by a double-covering rotation in the $SU(2)$, i.e. a corresponding transformation driven by the algebra of the $SU(2) \otimes SU(2)$, where $SU(2)$ is a subset of $SL(2, \mathbb{C})$, $SU(2) \subset  SL(2,\mathbb{C})$, which is homomorphic to the homogeneous Lorentz group $SO(1,3)$ (see Appendix).

The fundamental representation of the objects belonging to the homogeneous Lorentz group $SO(1,3)$, according to the homomorphism with $SU(2) \otimes SU(2)$, are the Dirac bispinors which can be described, for instance, either by the above-mentioned double-doublet representation of the {\em left-right} $SU(2)\otimes SU(2)$ group,
$ \psi^{\dagger}  =\left(
\psi _{\L\1},\,
\psi _{\L\2},\,
\psi _{\R\1},\,
\psi _{\R\2}\right)
\equiv (\bm{2},\bm{1})\oplus(\bm{1},\bm{2})$
or even by any other inequivalent representation of the $SU(2)\otimes SU(2)$.
That is the case of the solutions of the free particle Dirac equation put in the Lorentz covariant notation, $(\gamma^\mu p_\mu \mp m)u_{\pm,s}\bb{\bm{p}} = 0$,
\footnotesize\begin{equation}
\label{bispinors}
u_+\bb{{\bm{p}}} = \frac{1}{\sqrt{2 m (E_{p} + m)}}\left[ \begin{array}{rl} (E_{p} + m) &\chi_s \bb{\hat{\bm{p}}}\\ \bm{p} \cdot \bm{\sigma} &\chi_s \bb{\hat{\bm{p}}} \end{array}\right] \,\,~~ \mbox{and} ~~\,\, u_-\bb{{\bm{p}}} =\frac{1}{\sqrt{2 m (E_{p} + m)}} \left[ \begin{array}{rl} \bm{p} \cdot \bm{\sigma} &\chi_s \bb{\hat{\bm{p}}} \\ (E_{p} + m)& \chi_s \bb{\hat{\bm{p}}} \end{array}\right], 
\end{equation}\normalsize
with $p_{\mu} \equiv (E_p, \, -\bm{p})$, $\gamma^{\0} = \beta$ and $\gamma^i = \beta \alpha^i$, such that $\{\gamma^{\mu},\,\gamma^{\nu} \} = 2g^{\mu\nu}$, and which, for the matter of our proposal, are eigenstates of the free particle Dirac Hamiltonian Eq.~(\ref{twoqubithamiltonian}) identified by $\vert u_{\pm,s}\bb{\bm{p}}\rangle$, associated to positive and negative eigenvalues (associated frequencies), $\pm E_p = \pm\sqrt{p^2+m^2}$, which can be read as spin-parity entangled states given by
\begin{eqnarray}
\label{twoqubitspinor22}
\vert \, u_{\pm,s}\bb{\bm{p}} \, \rangle &=&  \frac{1}{\sqrt{2 m (E_{p} + m)}}\left[ (E_{p} + m)\,\, \vert \pm \rangle \otimes \vert \chi_s \bb{\bm{p}} \rangle \,\,+\,\, \vert \mp \rangle \otimes \,(\, \bm{p} \cdot \bm{\sigma} \, \vert \chi_s \bb{\bm{p}} \rangle )\right],\end{eqnarray}
corresponding to composite quantum systems in a total Hilbert space $\mathcal{H} = \mathcal{H}_P \otimes \mathcal{H}_S$ of spin polarization, $\vert \chi_s \bb{\bm{p}} \rangle$, and intrinsic parity, $\vert \pm \rangle$, quantum states, and for which the normalization properties shall be discussed in the following, in agreement with Lorentz covariance statements.

To clear up the meaning of the covariant behavior of the above structures under $SO(1,3)$ transformations, let one reconstructs the above states from some proper reference frame, i.e. where $\bm{p}$ is set equals to zero, such that the initial solutions of the Dirac equation are given by $u_{\pm,s}\bb{\bm{0}}$.

Firstly, one notices that the action of the linear transformation, $\Lambda$, describing a Lorentz homogeneous transformation results into the change of space-time coordinates between two inertial frames, $\mathcal{O}$ and $\mathcal{O}^{\prime}$, moving with respect to each other, such that the particle's quadrimomenta in each frame are related by $p^{\nu\prime} = \Lambda^{\nu}_{\mu} p^{\mu}$ ($\equiv \mt{p}^{\prime} = \Lambda \mt{p}$ in the matricial representative notation), where $p$ and $p^{\prime}$ follows from the respective notation for $\mathcal{O}$ and $\mathcal{O}^{\prime}$.
Once applied to spinors in the momentum representation, $w\bb{\mt{p}}$, the homogeneous Lorentz transformations result into 
\begin{equation}
\label{eq:12}
w\bb{\mt{p}} \to w^{\prime\hspace{-.4mm}}\bb{\mt{p}^{\prime\hspace{-.4mm}}} = U^{\mi\1}\bb{\Lambda} w\bb{\Lambda \mt{p}}\,U\bb{\Lambda}  = S\bb{\Lambda}w\bb{\mt{p}}, \end{equation}
where $S\bb{\Lambda}$ has to be obtained by following an overall adequacy to the covariant properties of the Dirac equation, which returns a simple consistency relation \cite{greiner}
\begin{equation}
S^{\mi\1} \bb{\Lambda} \, \gamma^{\mu}\, S\bb{\Lambda}\, \Lambda^{~\omega}_{\mu} = \gamma^\omega \qquad \leftrightarrow \qquad
S^{\mi\1} \bb{\Lambda}\, \gamma^{\mu}\, S \bb{\Lambda} = \Lambda^{\mu} _{~\zeta} \gamma^\zeta, 
\label{inversa}
\end{equation}
that leads to the explicit forms of $S\bb{\Lambda}$ decomposed into two subsets: one for $SO(1,3)$ Lorentz boosts and the other for $O(3)$ rotations.

The subset of Lorentz transformations corresponding to $SO(1,3)$ Lorentz boosts are given by
\begin{eqnarray}
S\bb{\Lambda} &=& \exp\left(i\, \eta\, \hat{\bm{p}}\cdot\mathbf{K}{\5}\right) =
\exp\left(\frac{\eta}{2}\, \gamma^{\5}\hat{\bm{p}}\cdot\mathbf{\Sigma}\right)=
\cosh{\left(\frac{\eta}{2}\right)} + \gamma^{\5}(\mathbf{\Sigma}\cdot \hat{\bm{p}})\sinh{\left(\frac{\eta}{2}\right)}\nonumber\\
&=&
 \frac{\gamma^\mu p_\mu\gamma^{\0} + m}{\left[2\, m\, (m+E)\right]^{\frac{\1}{\2}}},
\end{eqnarray}
where $\gamma^{\5}=i\gamma^{\0} \gamma^{\1} \gamma^{\2} \gamma^{\3}$, the vector components $K^j_{\5} = - (i/2)\, \gamma^{\5}\Sigma^j = (i/4)[\gamma^{j}, \gamma^{\0}]$ with $\Sigma^j = \gamma^{\5} \alpha^j$, and $\eta = \mbox{arcsinh}(\vert {\bm{p}}\vert /m)$ is the {\em rapidity} parameter related to the momentum $p$ boosted from the rest frame.

Once applied to $u_{\pm,s}\bb{\bm{0}}$, the covariant boost operator, $S\bb{\Lambda}$, leads to
\small\begin{eqnarray} 
S\bb{\Lambda} u_{\pm,s}\bb{\bm{0}} &=& \exp\left(i\, \eta\, \hat{\bm{p}}\cdot\mathbf{K}{\5}\right)\, u_{\pm,s}\bb{\bm{0}} = 
 \pm \frac{\gamma^\mu p_\mu + m}{\left[2\, m\, (m+E)\right]^{\frac{\1}{\2}}} \,u_{\pm,s}\bb{\bm{0}} = u_{\pm,s}\bb{\bm{p}},\label{403}
\end{eqnarray}\normalsize
from which, the  orthonormalization properties,
\begin{equation}
{u}^{\dagger}_{\pm,s}\bb{\bm{p}}\, {u}_{\mp,s}\bb{-\bm{p}} = 0,\qquad
{u}^{\dagger}_{\pm,s}\bb{\bm{p}}\, {u}_{\pm,r}\bb{\bm{p}} = (E_p/m) \delta_{sr},
\label{404BB}
\end{equation}
can be put into an equivalent Lorentz covariant form,
\begin{equation}
\overline{u}_{\pm,s}\bb{\bm{p}}\,{u}_{\mp,s}\bb{\bm{p}} = 0,\qquad
\overline{u}_{\pm,s}\bb{\bm{p}}\,{u}_{\pm,r}\bb{\bm{p}} = \pm \delta_{sr},
\label{40404}
\end{equation}
for which an auxiliary notation for $\overline{w} = w^{\dagger} \gamma^{\0}$ has been introduced, and the bi-linear Lorentz invariant product\footnote{Given that $\gamma^{\0} S^{\dagger}\gamma^{\0} = S^{\mi\1}$ , one notices that $\overline{w} \, w$ transforms as $\overline{w} \, w = w^{\dagger} \gamma^{\0} w \to w^{\dagger}\,S^{\dagger} \gamma^{\0} S \, w = w^{\dagger}\,  \gamma^{\0}S^{\mi\1} S \, w  =  w^{\dagger} \gamma^{\0} w = \overline{w} \, w$, which is therefore Lorentz invariant.}, $\overline{u}_{\pm,s}\bb{\bm{p}}\,{u}_{\pm,r}\bb{\bm{p}}$, as well as the covariant partial closure relations,
\begin{equation}
\pm\sum_{s={\1, \, \2}}{u}_{\pm,s}\bb{\bm{p}}\ \,\overline{u}_{\pm,s}\bb{\bm{p}}\ = \frac{1}{2}\left(1\pm \frac{\gamma^\mu p_\mu}{m}\right),
\label{4041A}
\end{equation}
have been identified.
 
Similarly, the subset of Lorentz transformations corresponding to $O(3)$ rotations are resumed by
\begin{eqnarray}
S\bb{\Lambda} &=& \exp\left(i\, \bm{\theta}\cdot\mathbf{\Sigma}\right) =
\cos{\left(\frac{\theta}{2}\right)} + i (\mathbf{\Sigma}\cdot \hat{\bm \theta})\sin{\left(\frac{\theta}{2}\right)}, 
\end{eqnarray}
with $\bm{\theta}=\theta\hat{\bm \theta}$, where the rotational axis $\hat{\bm \theta}$, for instance, can be identified by $\hat{\bm{p}}$.

The explicit expressions for $S\bb{\Lambda}$ reveals that, besides $S = S^{\dagger}$ for Lorentz boosts, one also has $S^{\mi\1} = S^{\dagger}$ for $O(3)$ rotations.
Given that $\gamma^{\0} S^{\dagger}\gamma^{\0} = S^{\mi\1}$ for boosts and rotations, one notices that $[S^{\dagger}, \gamma^{\0}] = [S, \gamma^{\0}] =0$ only for rotations.
Therefore, it is clearly relevant that the covariant behavior under $O(3)$ rotations is reflected into an independent (complete) invariance of the normalization conditions expressed either by 
Eqs.~\eqref{404BB} or by Eqs.~\eqref{40404}.
The same is not true for $SO(1,3)$ Lorentz boosts, for which only the set of Eqs.~\eqref{40404} are invariant.

Consequently, the choice of the metric normalization structure of the {\em spinorial} states drives the way of how the continuous Lorentz transformations do affect the probabilistic distribution of the quantum information content of such states, which is namely relevant when $SO(1,3)$ Lorentz boosts are discussed.
Through an optimized procedure involving the associated density matrix solutions related to Eqs.~(\ref{404BB}) and/or (\ref{40404}), quantum information quantifiers can thus be encompassed by a $SO(1,3)$ covariant description where quantum purity and quantum concurrence are typically Lorentz invariant quantities.

To clarify such an assertion, one firstly introduces a set of (non-covariant) projection operators at rest frame given by
\begin{equation}
\mathcal{P}_{\pm}^s = u_{\pm,s}\bb{\bm{0}} u_{\pm,s}^{\dagger}\bb{\bm{0}} = \frac{1}{4}\left(\mathbb{I}_4 + (-1)^{s\mi\1} \mathbf{\Sigma}\cdot \hat{\bm{k}} \right)\left(\mathbb{I}_4 \pm \gamma^{\0}\right),
\end{equation}
which, at rest frame, does not reflect any dependence on the choice of the normalization conditions, and where $\hat{\bm{k}} = \cos(\varphi)\sin(\vartheta)\hat{x} + \sin(\varphi)\sin(\vartheta)\hat{y} + \cos(\vartheta)\hat{z}$ is an arbitrary direction used to define $\chi_s\bb{\hat{\bm{k}}}$.

The projectors, $\mathcal{P}_{\pm}^s$, and their corresponding covariant modified versions, $\pm\mathcal{P}_{\pm}^s\gamma^{\0}$, are helpful in defining the boosted (or rotated) density matrix operators respectively consistent with the normalization condition Eqs.~(\ref{404BB}-\ref{40404}).
Starting from the free particle solutions at rest frame, in both cases, a trace unipotent property is identified by $Tr[(\mathcal{P}_{\pm}^s)^n] = Tr[(\pm\mathcal{P}_{\pm}^s\gamma^{\0})^n] = 1$.
Therefore, at rest frame, both hypothesis for defining density matrix operators associated to pure states $u_{\pm,s}\bb{\bm{0}}$, that are $\mathcal{P}_{\pm}^s$ and $\pm\mathcal{P}_{\pm}^s\gamma^{\0}$, are independently consistent with normalization conditions from Eqs.~(\ref{404BB}) and (\ref{40404}). In fact one has $\mathcal{P}_{\pm}^s = \pm\mathcal{P}_{\pm}^s\gamma^{\0}$, which turns the choice into a trivial problem.

However, considering that through the first hypothesis (cf. Eqs.~(\ref{404BB})) one identifies the density matrix operators ${\rho}_{\pm}^s$ with $\mathcal{P}_{\pm}^s$, which transform as $S\bb{\Lambda}\,\mathcal{P}_{\pm}^s \,S^{\dagger}\bb{\Lambda}$ under Lorentz boosts, the unipotent property of
$\mathcal{P}_{\pm}^s$ at rest frame is converted into a more enhanced problem which, at first glance, requires the introduction of a multiplicative factor $m/E_p = \cosh^{\mi\1}(\eta)$ as to preserve the probabilistic distribution.
Under Lorentz boosts, ${\rho}_{\pm}^s$, which transforms as $S\bb{\Lambda}{\rho}_{\pm}^s S^{\dagger}\bb{\Lambda}= S\bb{\Lambda}\mathcal{P}_{\pm}^s S^{\dagger}\bb{\Lambda}$, should be redefined as
\begin{equation}
{\rho}^{\prime s}_{\pm} = \cosh^{\mi\1}(\eta)\,S\bb{\Lambda}\mathcal{P}_{\pm}^s S^{\dagger}\bb{\Lambda}, 
\label{boosted}
\end{equation}
as to preserve the following trace properties,
\begin{equation}
Tr[{\rho}^{\prime s}_{\pm}] = \cosh^{\mi\1}(\eta)\,Tr[ S^{\dagger}\bb{\Lambda}\,S\bb{\Lambda}\,\mathcal{P}_{\pm}^s] = \cosh^{\mi\1}(0)\,Tr[{\rho}_{\pm}^s] = 1, 
\end{equation}
and the space-time local unitarity of the associate quantum theory\footnote{Even if it is already globally preserved by the Lorentz invariant continuity equation which sets
\begin{equation}
\partial_{\mu}(\overline{u}_{\pm}^s \gamma^{\mu}u_{\pm}^s) = 0.
\end{equation}}.

The introduction of the re-normalization factor, $\cosh(\eta)$, destroys completely the meaning of quantum purity from  $Tr[({\rho}^{\prime s}_{\pm})^{\2}]$ as well as of their now on non-unipotent versions $Tr[({\rho}^{\prime s}_{\pm})^{n}]$, which are not preserved under Lorentz boost (even if pure states remais as such).

The above aspects are all consistently circumvented by the second hypothesis suggested above, in which one assumes a strictly covariant form for density operators identified by $\bar{\rho}_{\pm}^s = \pm \mathcal{P}_{\pm}^s\,\gamma^{\0}$, which transform as
\begin{equation}
\label{TransformationLaw}
\bar{\rho}^{\prime s}_{\pm} = \pm\,S\bb{\Lambda}\,\mathcal{P}_{\pm}^s\,S^{\dagger}\bb{\Lambda}\, \gamma^{\0} = \pm S\bb{\Lambda}\,\mathcal{P}_{\pm}^s \,\gamma^{\0}\,S^{\mi\1}\bb{\Lambda} =S\bb{\Lambda}\,\bar{\rho}_{\pm}^s \, S^{\mi\1}\bb{\Lambda}, 
\end{equation}
such that 
\small\begin{equation}
\label{beber}
Tr[(\bar{\rho}^{\prime s}_{\pm})^n] = (\pm 1)^{n} Tr[S\bb{\Lambda}\,\mathcal{P}_{\pm}^s \,\gamma^{\0}\,S^{\mi\1}\bb{\Lambda}S\bb{\Lambda}\,\dots\, S^{\mi\1}\bb{\Lambda}S\bb{\Lambda}\,\mathcal{P}_{\pm}^s \,\gamma^{\0}\,S^{\mi\1}\bb{\Lambda}] = Tr[(\bar{\rho}_{\pm}^s)^n] = 1,
\end{equation}\normalsize
that corresponds to a unipotent invariant quantity.

Of course, given that $S^{\dagger}\bb{\Lambda} = S^{\mi\1}\bb{\Lambda}$ for Lorentz transformations corresponding to $O(3)$ rotations, for both hypothesis discussed above, in terms of $\bar{\rho}_{\pm}^s$ and of ${\rho}_{\pm}^s$, the trace unipotent properties are straightforwardly preserved (i.e. $Tr[(\bar{\rho}^{\prime s}_{\pm})^n] =Tr[({\rho}^{\prime s}_{\pm})^n] = 1$ for $O(3)$ rotations).
Therefore, it is worth to mention that \cite{Caban} the {\em boosted} operators from Eq.~(\ref{boosted}) exhibit a nonlinear realization of the Lorentz group connected with the quotient space $SO(1,3)_o/SO(3)$, which is linear on the rotation group. It means that extracting information about polarization states is not affected by the above choices.
Otherwise, the intrinsic entanglement measured by the quantum concurrence between polarization and intrinsic parity is straightforwardly affected by the choice of the normalization conditions and by the respective density operator, as it shall be discussed in the following section.

\section{Invariance of Quantum Concurrence} 

The entanglement, seen as a consequence of the paradigm that rules the quantum superposition, is properly defined through the concept of separability: entangled states are not separable.
To clarify the above statement, one assumes that a quantum state in a so-called bi-partite configuration is separable if it is possible to express its associated density operator $\varrho$ as \cite{n021}
\begin{equation}
\varrho = \sum_{i} q_i \, \sigma_i ^{(1)} \otimes \tau_i ^{(2)},
\end{equation}
with $\sigma_i ^{(1)} \in \mathcal{H}_1$, $\tau_i ^{(2)} \in \mathcal{H}_2$ and $\sum_i q_i = 1$. 

When density operators are of the form $\vert w \rangle \langle w \vert$, which describes pure states, the separability issue can be posed into the state vector notation: $\vert w \rangle = \vert w_1 \rangle \otimes \vert w_2 \rangle$, for instance, are separable states. 
Entanglement is quantified through different methods, depending on the context and on the specific properties that are considered \cite{n022}. 
More generically, for pure states, the Schmidt decomposition theorem establishes that the traced-out density operators $\varrho_{1 \, (2)} = Tr_{2 (1)}[\varrho]$ have equal eigenvalues. If the state is entangled then either $\varrho_{1(2)}$ are mixed states. 
In this case, the quantum correlation between 1 and 2 Hilbert spaces can be simply described by the entanglement entropy $E_{vN}[\varrho]$ given by
\begin{equation}
E_{vN}[\varrho] = S[\varrho_2] = - Tr_2[\varrho_2 \log_2 \varrho_2] = S[\varrho_1] = - Tr_1[\varrho_1 \log_2 \varrho_1],
\end{equation}
where $S[\varrho]$ is the von Neumann entropy of the involved susbystems \cite{n021}.
The entanglement can indeed be quantified through several theoretical tools \cite{n022}. 
For instance, in case of mixed states, entanglement quantifiers can be built by means of the convex-roof extension of pure state entanglement quantifiers \cite{n023}.
In the scope of the formulation of a Lorentz covariant theory for quantum correlations, in the particular context of Dirac spinor quantum states, quantum entanglement for a generic quantum state described by $\varrho$, the entanglement of formation \cite{n024} -- the convex-roof extension of $E_{vN}[\varrho]$ -- defined as the  average of the pure-state entanglement, minimized over all decompositions of the mixed state $\varrho$ on pure states, $\varrho_k$,
\begin{equation}
E_{EoF} [\varrho] = \mbox{min}_{\varrho_k} \displaystyle \sum_k q_k E_{vN}[\varrho_k],
\end{equation}
is a prominent candidate for observing Lorentz invariance properties.

For two-qubit states, the entanglement of formation is given by
\begin{eqnarray}
E_{EoF} [\varrho] &=& \mathcal{E}\left[ \frac{1 - \sqrt{1 - \mathcal{C}^2[\varrho]}}{2}\right],
\end{eqnarray}
with $\mathcal{E}[\lambda] =- \lambda \log_2 \lambda - (1-\lambda)\log_2 (1-\lambda)$, and where the quantum concurrence, $\mathcal{C}[\varrho]$, by itself is an entanglement quantifier defined as \cite{n024}
\begin{equation}
\mathcal{C}[\varrho] = \mbox{max}\{ \omega_1 - \omega_2 - \omega_3 - \omega_4 \, , \,0 \},
\end{equation}
where $\omega_1 > \omega_2 > \omega_3 > \omega_4$ are the eigenvalues of the operator $\sqrt{\, \sqrt{\varrho} \, (\sigma_y \otimes \sigma_y) \varrho^\ast (\sigma_y \otimes \sigma_y) \, \sqrt{\varrho}\, }$. 

The point here is that, for two-qubit quantum pure states, a highly simplified expression for quantum concurrence can be given in terms of a qubit-flip operation \cite{n024,Osb05}, which corresponds to an antilinear operation.
To be more specific, for a generic pure state described by $\vert w \rangle\langle w\vert$, the qubit-flip operation returns  $\vert \tilde{w} \rangle\langle \tilde{w}\vert$, with
\begin{equation}
\vert \tilde{w} \rangle = \sigma^{(1)}_y\otimes\sigma^{(2)}_y\vert {w}^* \rangle,
\end{equation}
where ``$*$'' denotes the complex conjugation operator.
For pure states, $\varrho = \vert w \rangle\langle w\vert$, the quantum concurrence is thus expressed by
\begin{equation}
\mathcal{C}[\varrho] = \sqrt{\langle w  \vert \tilde{\rho}\vert w\rangle} = \vert \langle w \vert \tilde{w} \rangle\vert = \sqrt{Tr [ \varrho \tilde{\varrho}]},
\end{equation}
with $\tilde{\varrho} = \vert \tilde{w} \rangle \langle \tilde{w} \vert$.
For statistical mixtures of two-qubits, otherwise, the quantum concurrence is defined as the minimum over all pure-state decompositions $\{q_i, w_i\}$ of $\varrho$ \cite{Osb05},
\begin{equation}
\mathcal{C}[\varrho] = \min_{\{q_i, w_i\}} \sum_i q_i\vert \langle w_i \vert \tilde{w}_i \rangle\vert.
\end{equation}

If a quantum state $\varrho$ is generically identified by
\begin{equation}
\label{D04}
\varrho = \frac{1}{4} \left[ I + (\bm{\sigma}^{(1)} \otimes I^{(2)}) \cdot \bm{a} + (I^{(1)} \otimes \bm{\sigma}^{(2)}) \cdot \bm{b}  + \displaystyle \sum_{i,j = 1}^3 t_{ij} (\sigma_i^{(1)} \otimes \sigma_j^{(2)}) \right],
\end{equation}
where $t_{ij}$ are the elements of the correlation matrix, $\bm{T}$, and $\bm{a}$ and $\bm{b}$ are the Bloch vectors of the corresponding subsystems, it is straightforward to notice that, for pure states, $a^2 = b^2$ and the concurrence is given by
\begin{equation}
\mathcal{C}[\varrho] = \sqrt{1 - a^2}.
\end{equation}

Turning back to the aspects of the Lorentz covariant structure of spinor quantum states, an intrinsic quantum correlation between spin and parity can be evaluated in terms of the quantum concurrence of ${\rho}_{\pm}^s$ and of $\bar{\rho}_{\pm}^s$.
Once that the qubit-flip operator $\sigma^{(P)}_y\otimes\sigma^{(S)}_y$ is identified with $-i\,\gamma^{\2}$, the corresponding qubit-flipped density matrices are given by
\begin{equation}
\tilde{\rho}_{\pm}^s =(-i\,\gamma^{\2}) \,u^*_{\pm,s} u_{\pm,s}^{T} \,(-i\,\gamma^{\2})
\end{equation}
and
\begin{equation}
\tilde{\bar{\rho}}_{\pm}^s =\pm (-i\,\gamma^{\2}) \,u^*_{\pm,s} u_{\pm,s}^{T} \,(-i\,\gamma^{\2})\gamma^{\0}
\end{equation}
where ``$^T$'' is the transposed matrix, such that
\small\begin{equation}
Tr[\tilde{\rho}_{\pm}^s] =Tr[(-i\,\gamma^{\2}) \,u^*_{\pm,s} u_{\pm,s}^{T} \,(-i\,\gamma^{\2})] = Tr[-(\gamma^{\2})^2 \,(u_{\pm,s} u_{\pm,s}^{\dagger})^T] = Tr[u_{\pm,s} u_{\pm,s}^{\dagger}] = Tr[{\rho}_{\pm}^s],
\end{equation}\normalsize
as well as $Tr[\tilde{\bar{\rho}}_{\pm}^s] = Tr[\bar{\rho}_{\pm}^s]$, where it has been noticed that $(\gamma^{\2})^2 = -1$.
Considering the calculation of the square of the quantum concurrence under the action of a continuous Lorentz transformation, $S\bb{\Lambda}$, one typically obtains
\begin{equation}
\mathcal{C}^2[{\rho}_{\pm}^s] = Tr[{\rho}_{\pm}^s \tilde{\rho}_{\pm}^s] \stackrel{S(\Lambda) }{\rightarrow} 
Tr[(S {\rho}_{\pm}^s S^{\dagger}) ((-i\gamma^{\2}) S^* {\rho}_{\pm}^{s*} S^T (-i\gamma^{\2}))] = Tr[S^{\dagger}S\,{\rho}_{\pm}^s\, S^{\dagger}S\, \tilde{\rho}_{\pm}^s],
\end{equation}
for the first hypothesis, where it has been noticed that $(-i\gamma^{\2}) S = S^*(-i\gamma^{\2})$ and $(-i\gamma^{\2}) S^{\dagger} = S^T(-i\gamma^{\2})$, and which is not invariant under Lorentz boosts, in spite of being invariant under $O(3)$ rotations. Analogously, by following the covariant hypothesis for $\bar{\rho}_{\pm}^s$, one obtains
\begin{eqnarray}
\mathcal{C}^2[\bar{\rho}_{\pm}^s] = Tr[\bar{\rho}_{\pm}^s \tilde{\bar{\rho}}_{\pm}^s] &\stackrel{S(\Lambda) }{\rightarrow}&  
Tr[(S \bar{\rho}_{\pm}^s S^{\mi\1}) ((-i\gamma^{\2}) (S (\pm 1){\rho}_{\pm}^s S^{\dagger})^* \gamma^{\0*}  (-i\gamma^{\2}))]\nonumber\\&=& Tr[(S \bar{\rho}_{\pm}^s S^{\mi\1}) ((-i\gamma^{\2}) S^*) (\pm 1){\rho}_{\pm}^{s*} S^{T}  (+i\gamma^{\2})\gamma^{\0}] \nonumber\\&=& 
Tr[S \bar{\rho}_{\pm}^s (S^{\mi\1}S)(-i\gamma^{\2}) (\pm 1){\rho}_{\pm}^{s*} (+i\gamma^{\2})S^{\dagger}\gamma^{\0}] \nonumber\\&=&
Tr[S \bar{\rho}_{\pm}^s (-i\gamma^{\2}) (\pm 1){\rho}_{\pm}^{s*} (+i\gamma^{\2})\gamma^{\0} S^{\mi\1}] \nonumber\\&=&[ \bar{\rho}_{\pm}^s (-i\gamma^{\2}) (\pm 1){\rho}_{\pm}^{s*} \gamma^{\0} (-i\gamma^{\2})] \nonumber\\&=& Tr[\bar{\rho}_{\pm}^s \tilde{\bar{\rho}}_{\pm}^s],
\end{eqnarray}
which is a totally invariant quantity under continuous Lorentz transformations.

In order to confront the above results, let one considers a typical Dirac Hamiltonian with a magnetic field interacting term for a Dirac particle, 
\begin{equation}
\label{Btwoqubithamiltonian}
{H} = \gamma^{\0}(m +\mu \mathbf{\Sigma}\cdot {\bm{B}}) \equiv m ( {\sigma}_z ^{(P)} \otimes {I}_{2}^{(S)}) +\mu  ( {\sigma}_z ^{(P)} \otimes {\bm{\sigma}}^{(S)}\cdot {\bm{B}}),
\end{equation}
from which, both choices for the normalization scheme lead to pure (eigen)state density matrices given by
\begin{eqnarray}
{\rho}_{\pm}^{s,B} &=& \frac{1}{4}\left(1 + (-1)^{s\mi\1} \mathbf{\Sigma}\cdot {\hat{\bm{B}}}\right)
\left(1 \pm \gamma^{\0} \frac{m + \mu \mathbf{\Sigma}\cdot {\bm{B}}}{m+(-1)^{s\mi\1}\mu B}\right)\nonumber\\
&=& \frac{1}{4}\left(1 \pm \gamma^{\0}\right) \left(1 + (-1)^{s\mi\1}\mathbf{\Sigma}\cdot {\hat{\bm{B}}}\right),
\end{eqnarray}
which coincidently has the same form of $\bar{\rho}_{\pm}^{s,B}$.
In particular, it straightforwardly results into the eigenvalues given by
\begin{equation}
Tr[{\rho}_{\pm}^{s,B} \, H ] =  Tr[\bar{\rho}_{\pm}^{s,B} \, H ]= \pm\left(m+(-1)^{s\mi\1}\mu B\right).
\end{equation}
The boosted form of the above quantum states are given by 
\begin{eqnarray}
{\rho}_{\pm}^{s,B}\bb{\bm{p}} &=& \frac{m}{E_p} S\, {\rho}_{\pm}^{s,B} \, S^{\dagger}\nonumber\\ 
&=&\frac{1}{4}\left\{1 +\gamma^{\5}\frac{\bm{\Sigma}\cdot {\bm{p}}}{E_p}\pm \gamma^{\0}
\frac{m}{E_p}
\right.\nonumber\\&&\qquad  \left.+ (-1)^{s\mi\1}\left[\left(\frac{m}{E_p}\pm \gamma^{\0}\right)\mathbf{\Sigma}\cdot {\hat{\bm{B}}} +
\frac{\bm{p}\cdot {\hat{\bm{B}}}}{E_p}\left(\gamma^{\5} +(1\mp\gamma^{\0})\frac{\bm{\Sigma}\cdot {{\bm{p}}}}{E_p+m} \right)\right]\right\},
\end{eqnarray}
with $S \equiv S\bb{\Lambda\bb{\bm{p}}}$, and by
\begin{eqnarray}
\bar{\rho}_{\pm}^{s,B}\bb{\bm{p}} &=&  S\, \bar{\rho}_{\pm}^{s,B} \, S^{\mi\1}\nonumber\\ &=&
\frac{1}{4}\left\{1 \pm \gamma^{\0}\left(\frac{E_p -\gamma^{\5}\bm{\Sigma}\cdot {\bm{p}}}{m}\right)
\right.\nonumber\\&&\qquad  \left.+ (-1)^{s\mi\1}\left[\left(\frac{E_p}{m}\pm \gamma^{\0}\right)\mathbf{\Sigma}\cdot {\hat{\bm{B}}} -
\frac{\bm{p}\cdot {\hat{\bm{B}}}}{m}\left((1\mp\gamma^{\0})\frac{\bm{\Sigma}\cdot {{\bm{p}}}}{E_p+m} \pm\gamma^{\0}\gamma^{\5} \right)\right]\right\},
\end{eqnarray}
respectively for the re-normalized usual quantum mechanical picture and for the covariant picture, and where the term proportional to $\mathbf{\Sigma}\cdot {(\bm{p}\times\hat{\bm{B}})}$ was suppressed as to simplify the role of the boosted magnetic field, given that $\bm{B}$ remains unchanged along the direction of the boost.
Correspondently, the Bloch vectors are given by $\bm{a} = E_p^{\mi\1}(m,\,0,\,\bm{p}\cdot {\hat{\bm{B}}})$ and $\bm{b} = E_p^{\mi\1}(m\hat{\bm{B}}+ \bm{p}(\bm{p}\cdot {\hat{\bm{B}}})/(E_p+m))$ for ${\rho}_{\pm}^{s,B}\bb{\bm{p}}$, and by $\bar{\bm{a}} = m^{\mi\1}(E_p,\,i \bm{p}\cdot {\hat{\bm{B}}},\,0)$ and $\bar{\bm{b}} = m^{\mi\1}(E_p\hat{\bm{B}}- \bm{p}(\bm{p}\cdot {\hat{\bm{B}}})/(E_p+m))$ for $\bar{\rho}_{\pm}^{s,B}\bb{\bm{p}}$.
In this case, one obtains that 
$$\mathcal{C}^2[{\rho}_{\pm}^{s,B}\bb{\bm{p}}] =\sqrt{1-({a}^2+{b}^2)/2} =
\dots \neq 0,$$
 and that $$\mathcal{C}^2[\bar{\rho}_{\pm}^{s,B}\bb{\bm{p}}] = \sqrt{1-(\bar{a}^2+\bar{b}^2)/2} = 0,$$
given that $\bar{a}^2 = \bar{b}^2 = (E_p^2 - (\bm{p}\cdot {\hat{\bm{B}}})^2)/ m^2=1$ is a Lorentz invariant quantity.

In certain sense, the above result just exemplifies the already known incompatibility between Lorentz invariance and Hermitian quantum mechanics. Despite the non-Hermitian characteristic of the boosted density matrix operator, $\bar{\rho}_{\pm}^{s,B}\bb{\bm{p}}$, physical observables and measurement (trace) properties are all preserved and do not violate the probabilistic and statistic axioms of quantum mechanics \cite{Str64}. For instance, similar problems involving the normalization metric definition, related to such a framework incompatibility, are identified along the Gupta-Bleuler Lorentz invariant procedure for the second quantization of electromagnetic fields \cite{Gupta,Bleuler}. Even if the Hermitian version, ${\rho}_{\pm}^{s,B}\bb{\bm{p}}$, can be recovered from $\bar{\rho}_{\pm}^{s,B}\bb{\bm{p}}$, in the former case, the Lorentz boost preserves Hermicity and positive semi-definiteness of ${\rho}_{\pm}^{s,B}\bb{\bm{p}}$, but changes its trace and, therefore, does not preserve quantum correlation observables. Of course, all the above mentioned points are encompassed by a deeper analysis of the compatibility between quantum mechanics and special relativity.

Turning ones attention to a still partially open problem which demands for some strategy to quantify the entanglement of mixed states (that can eventually exhibit quantum correlations other than entanglement \cite{n025}), it is worth noting that the demonstrated Lorentz invariance of $\mathcal{C}[\bar{\rho}]$ can be used as an important component tool for obtaining Lorentz invariant entanglement quantifiers for rank 2 mixed states, which are indeed exactly measurable \cite{Osb05,Yu08}. For a two-qubit state, the concurrence of a given rank 2 mixed state $\varrho$ is given by \cite{Hill97}
\begin{equation}
\mathcal{C}[\varrho] = \vert \omega_1 - \omega_2\vert,
\end{equation}
where $\omega_i$ are the square roots of the eigenvalues of $\varrho\tilde{\varrho}$ in decreasing order.
It is possible to demonstrate that \cite{Yu08} 
\begin{equation}
\mathcal{C}^2[\varrho] = Tr[\varrho\tilde{\varrho}] - \sqrt{2\left\{ \left(Tr[\varrho\tilde{\varrho}]\right)^2 - Tr[\left(\varrho\tilde{\varrho}\right)^2]\right\}}.
\end{equation}
Once $\varrho$ is identified with the covariant form of $\bar{\rho}$, one also has $\mathcal{C}^2[\bar{\rho}]$ as a Lorentz invariant quantity. As another example, several rank 2 mixed states can be obtained from our preliminary example, for instance, either as $\bar{\rho}^{s,B}\bb{\bm{p},\,q} = q \bar{\rho}^{s,B}_{+}\bb{\bm{p}} + (1-q)\bar{\rho}^{s,B}_{-}\bb{\bm{p}}$ or as $\bar{\rho}^{B}_{\pm}\bb{\bm{p},\,q} =q \bar{\rho}^{1,B}_{\pm}\bb{\bm{p}} + (1-q)\bar{\rho}^{2,B}_{\pm}\bb{\bm{p}}$, with $0 < q < 1$.
For exact helicity and parity projections, obtained by setting $q=1/2$ respectively for $\bar{\rho}^{s,B}$ and for $\bar{\rho}^{B}_{\pm}$ one has
\begin{equation}
\bar{\rho}^{s,B}\bb{\bm{p},\,1/2}  =
\frac{1}{4}\left\{1
+ (-1)^{s\mi\1}\left[\frac{E_p}{m}\mathbf{\Sigma}\cdot {\hat{\bm{B}}} -
\frac{\bm{p}\cdot {\hat{\bm{B}}}}{m}\left(\frac{\bm{\Sigma}\cdot {{\bm{p}}}}{E_p+m} \right)\right]\right\},
\end{equation}
and
\begin{equation}
\bar{\rho}^{B}_{\pm}\bb{\bm{p},\,1/2} = \frac{1}{4}\left\{1 \pm \gamma^{\0}\left(\frac{E_p -\gamma^{\5}\bm{\Sigma}\cdot {\bm{p}}}{m}\right)\right\},
\end{equation}
and therefore
$\mathcal{C}^2[\bar{\rho}^{s,B}\bb{\bm{p},\,1/2}]  = \mathcal{C}^2[\bar{\rho}^{B}_{\pm}\bb{\bm{p},\,1/2}] = 0$, since again all the traces involved in the calculation of $\mathcal{C}^2$ are functions of $(E_p^2 - (\bm{p}\cdot {\hat{\bm{B}}})^2)/ m^2=1$, that results into the preservation of the same values of $\mathcal{C}^2$ obtained at rest frame.
Correspondently, for $\bar{\rho}$ replaced by ${\rho}$, one has a complete distortion of any invariant pattern for $\mathcal{C}^2$ under Lorentz boosts.

Just to end up, as discussed in Refs.~\cite{extfields,SU2}, the spin-parity intrinsic structure of Dirac spinors supports the inclusion of Hamiltonian interacting properties as they appear in a full Dirac Hamiltonian like \cite{diraclike01,diraclike02}
\begin{eqnarray}
\label{E04T}
\hat{H}  &=& A^0\bb{\bm{x}}\,\hat{I}_4+ \hat{\beta}( m + \phi_S \bb{\bm{x}} ) + \hat{\bm{\alpha}} \cdot (\hat{\bm{p}} - \bm{A}\bb{\bm{x}}) + i \hat{\beta} \hat{\gamma}_5 \mu\bb{\bm{x}} - \hat{\gamma}_5 q\bb{\bm{x}} + \hat{\gamma}_5 \hat{\bm{\alpha}}\cdot\bm{W}\bb{\bm{x}} \nonumber \\
&+& i \hat{\bm{\gamma}} \cdot [ \zeta_a \bm{B}\bb{\bm{x}} + \kappa_a\, \bm{E}\bb{\bm{x}}  \,] + \hat{\gamma}_5 \hat{\bm{\gamma}}\cdot[\kappa_a\, \bm{B}\bb{\bm{x}}  - \zeta_a \bm{E}\bb{\bm{x}} \,],
\end{eqnarray}
which also transforms according to Poincar\'e symmetries described by the extended Poincar\'e group \cite{WuTung}.The above Hamiltonian admits the inclusion of interactions with an external vector field with time component $A^0\bb{\bm{x}}$ and spatial components $\bm{A} \bb{\bm{x}}$, a non-minimal coupling with magnetic and electric fields, $\bm{B}\bb{\bm{x}}$ and $\bm{E}\bb{\bm{x}}$ (via $\kappa_a$ and  $\zeta_a$) and also pseudovector field interactions with time component $q\bb{\bm{x}}$, and spatial components $\bm{W}\bb{\bm{x}}$, besides scalar and pseudoscalar field interactions through $\phi_S\bb{\bm{x}}$  and $\mu\bb{\bm{x}}$, respectively.
Algebraic strategies \cite{extfields} for obtaining the analytical expression for the matrix density of the associated eigenstates of the above Hamiltonian problem have been developed \cite{SU2,diraclike01,diraclike02}, however, they are out of the central scope of this paper.

\section{Conclusions}

Although the setup for describing the properties of quantum entanglement under Lorentz boost has suggested some elegant insights into the physics of quantum information, when massive charged spinor fields, like Dirac electrons, are considered, a completely covariant formulation of their related quantum information issues is still missing.
Electrons and any QED spinor interacting particles are indeed described by four component objects in a double-doublet   representation of the $SL(2,\mathbb{C})\otimes SL(2,\mathbb{C})$ group -- the Dirac bispinors, which satisfy the Dirac equation.
Therefore, a relativistically covariant formulation of quantum concurrence quantifiers -- as the first step into the direction for taking into account the subtle properties of massive spinors -- has been described in this work.

By following the relativistic quantum mechanics framework supported by Dirac bispinor symmetry properties, and providing a consistent covariant formulation of the probabilistic quantum mechanics paradigms, it has been shown that quantum concurrence of $SU(2) \otimes SU(2)$ entangled structures are invariant under homogeneous Lorentz transformations for quantum states constructed as spin-parity density matrices.
The Dirac structure of the covariant quantum mechanics formulation is reflected onto the covariant form of quantum purity and of the trace of unipotent density matrix operators, which are invariant under $SO(1,3)$ Lorentz boosts and $O(3)$ rotations. 
Consequently, the same $SU(2) \otimes SU(2)$ Dirac bispinor structure exhibits an entaglement profile that is invariant under Lorentz boosts and rotations, quantitatively preserved at any reference frame.

In particular, the parity symmetry is intrinsically involved into the above problem, given that parity operations exchange two {\em irreps} of the Poincar\'{e} group, such that a proper formulation must be engendered in terms of {\em irreps} of the complete Lorentz group.
The point here is that both the information about spin polarization and the $SL(2,\mathbb{C})\otimes SL(2,\mathbb{C})$ correlated information about intrinsic parity have been mutually involved into Dirac bispinor correlations.
It is demanded by the fact that $SU(2) \otimes SU(2)$ symmetry, is just one of the inequivalent representations enclosed by the $SL(2,\mathbb{C})\otimes SL(2,\mathbb{C})$ symmetry.
The related result for spin-parity quantum concurrence, and the quantum concurrence of any physical system which can be covered by the $SL(2,\mathbb{C})\otimes SL(2,\mathbb{C})$ symmetry where the $SU(2) \otimes SU(2)$ is homomorphic to $SO(1,3)$, thus achieve a more universal meaning where entanglement properties do not depend on the kinematics of the quantum systems.

The resulting covariant framework was used for computing the Lorentz invariant spin-parity quantum entanglement of spinorial particles coupled to a magnetic field, through which the extensions to more general Poincar\'e classes of spinor interactions can be straightforwardly investigated.

For this purpose, the subsidiary geometrical structure of Dirac spinors could be helpful in improving the understanding of Lorentz invariant entanglement properties for $SU(2) \otimes SU(2)$ coupled states.
The ground steps into this direction could also be noticed from Ref. \cite{cor1}, from which a standard Bloch sphere representation for a single qubit is generalized to two qubits, in the framework of Hopf fibrations of high-dimensional spheres.
In this case, a single-qubit Hilbert space is mapped onto a three-dimensional sphere $S^3$, and for a double-qubit, the Hilbert space is a seven-dimensional sphere $S^7$, from which, the striking result is that suitably oriented $S^7$ Hopf fibrations are entanglement sensitive \cite{cor1}.
Furthermore, subsequent steps in interpreting entanglement in geometric terms through a convenient quaternionic representation suggests that the two-qubit Hilbert space is fibred over the $S^4$ sphere, isomorphic to the one dimensional quaternionic projective space \cite{cor2}: the relationship between base and fibre supports the concept of qubit separability \cite{cor2}.
In general sense, the correspondence between Hopf fibrations and Dirac-like spinor structures suggests some equivalence in describing the intrinsic entanglement properties, which certainly deserves further investigations.

From a different perspective, but in a similar context, several elegant geometric formulations, still more enhanced with the subjacent $SL(2,\mathbb{C})$ Dirac spinor structure, have been considered in describing covariant quantum entanglement properties \cite{cor3,cor4}. Through the correspondence between 
two-qubit system with four-component Dirac spinors, the quantum entanglement using the mathematical formulation of Cartan's pure spinors has been worked out \cite{cor3}. Essentially, it has been shown how quantifiable intrinsic quantum correlations structures are exhibited by higher dimensional spinors engendered from elementary unitary representations of inhomogeneous $SL(2,\mathbb{C})$, which naturally are the ground structures for the spin-parity quantum correlations here discussed (see the Appendix).

To end up, considering that effects of Lorentz boosts have been also investigated in a plethora of quantum mechanical problems \cite{spins, celeri,bispinorarxiv02}, all the above-mentioned frameworks can be connected to our results as to stimulate one to hypothesize that a completely covariant formulation of quantum correlation issues can also be developed in the context of the scrutinization of Bell inequalities \cite{bellDirac} and, for instance, in order to extend the results for Fouldy-Wouthuysen (FW) spin operators \cite{greiner}, as to describe the transformation of spin-spin entanglement properties encoded in FW eigenstates \cite{bispinorFW}.

{\em Acknowledgments - The work of AEB is supported by the Brazilian Agencies FAPESP (grant 2018/03960-9) and CNPq (grant 301000/2019-0). VASVB acknowledge the support from the Max Planck Gesellschaft through an independent Max Planck Research Group.} 

\section*{Appendix -- {The group $SU(2)\otimes SU(2)$ as a subgroup of $SL(2,\mathbb{C})\otimes SL(2,\mathbb{C})$ and the \em spin-parity} intrinsic structure of Dirac bispinors}

\hspace{1 em} The group representations of $sl(2,\mathbb{C})\oplus sl(2,\mathbb{C})$ -- the Lie algebra of the $SL(2,\mathbb{C})\otimes SL(2,\mathbb{C})$ Lie group -- are irreducible, since they are tensor products of linear complex representations of $sl(2,\mathbb{C})$. Given that $SU(2)\otimes SU(2) \subset SL(2,\mathbb{C})\otimes SL(2,\mathbb{C})$, the unitary irreducible representations ({\em irreps}) of the $SU(2)\otimes SU(2)$ are built through tensor products between unitary representations of $SU(2)$. Such an one-to-one correspondence with the $SL(2,\mathbb{C})\otimes SL(2,\mathbb{C})$ simply connected group imposes a single correspondence with the $sl(2,\mathbb{C})\oplus sl(2,\mathbb{C})$ algebra.
In fact, the existence of {\em inequivalent representations} of $SU(2) \otimes SU(2)$ follows exactly from such one-to-one correspondences. 
The point is that inequivalent representations (for instance, the one for the chiral basis and another one for the spin-parity basis) do not correspond to the complete set of representations of $SL(2,\mathbb{C})\otimes SL(2,\mathbb{C})$, and therefore, each of them, independently, does not exhibit a one-to-one correspondence with the homogeneous Lorentz transformations that compose the algebra of the $SO(1,3)$ group.
The chiral basis described by a double-doublet representation of $SU(2)\otimes SU(2)$, for instance, maps a subset of transformations of the $SO(4) \equiv SO(3)\otimes SO(3)$ group, as for instance, those which include the double covering rotations.

Turning back to our point, since the $SU(2)\otimes SU(2)$ transformations can be mapped into a subset of $SL(2,\mathbb{C})\otimes SL(2,\mathbb{C})$, one may choose two {\em inequivalent} subsets of $SU(2)$ generators, such that $SU(2)\otimes SU(2) \subset SL(2,\mathbb{C})\otimes SL(2,\mathbb{C})$, with each group transformation generator having its own {\em irrep}.
In an overall context, despite the effectiveness of the {\em irreps} of the Poincar\'{e} group, in the Lorentz covariant Hamiltonian formulation of quantum mechanics, one has to pay attention to the inclusion of masses in the relativistic formalism described by the Dirac equation.
It requires the inclusion of the parity symmetry and the equalization of its role with the helicity (spin one-half projection, $\hat{\bm{e}}_p\cdot\hat{\boldmath{\mbox{$\sigma$}}} \sim \hat{\sigma}_z$) symmetry, as an accomplished $SU(2)$ symmetry.
Under such conditions, one designates the fundamental representation of the $SU_{\xi}(2)$ as a spinor-like object $\xi$ described by $(\pm,\,0)$, which transforms as a $SU_{\xi}(2)$ {\em doublet} ($2$-dim) of parity quantum numbers, $\pm$, and as a singlet ($1$-dim), which is transparent under any $SU_{\chi}(2)$ transformation.
Reciprocally, the fundamental object of the $SU_{\chi}(2)$, a typical spinor $\chi$ described by $(0,{1}/{2})$, transforms as a $SU_{\chi}(2)$ {\em doublet} of spin quantum numbers, $\pm 1/2$, and as a singlet of the $SU_{\xi}$\footnote{The {\em spinor} $\xi$ is an object dimensionally represented by given by $(\bm{dim}(SU_{\xi}(2)),\bm{dim}(SU_{\chi}(2)))$ =(\bm{2},\bm{1}).}.

The meaning of the above mentioned quantities related to parity and helicity can be clarified by noticing that the total parity operator $\hat{P}$ acts on the direct product $\left\vert \pm \right\rangle \otimes \left\vert \chi_s\bb{\bm{p}}\right\rangle$ as $$\hat{P}\left( \left\vert \pm \right\rangle \otimes \left\vert \chi_s\bb{\bm{p}}\right\rangle\right) =\pm \left( \left\vert \pm \right\rangle\otimes \left\vert \chi_s(-\bm{p})\right\rangle\right),$$ and, for instance, it corresponds to the Kronecker product of two operators, $\hat{P}^{(P)}\otimes \hat{P}^{(S)}$, where $\hat{P}^{(P)}$ is the intrinsic parity (with two eigenvalues, $\hat{P}^{(P)}\left\vert \pm \right\rangle =\pm \left\vert\pm \right\rangle $) and $\hat{P}^{(S)}$ is the spatial parity (with $\hat{P}^{(S)}\chi_s \left( \bm{p}\right) =\chi_s \left( -\bm{p}\right) $).

The above construction supports the identification of Dirac Hamiltonian eigenstates with electrons in the double-doublet {\em irrep} of the $SU(2) \otimes SU(2)$. Spatial parity couples quantum states with $\pm$ parity and $\pm 1/2$ spin quantum numbers given that they are described by {\em irreps} of the Poincar\'{e} group \cite{fonda}. To comprehend the covariant behavior of parity, one needs to consider the {\em extended Poincar\'{e} group} \cite{weinberg,WuTung} which also accounts for the helicity (spin-projection) in a double-doublet $SL(2,\mathbb{C}) \otimes SL(2,\mathbb{C})$ representation given by Dirac four component spinors, the bispinors satisfying the Dirac equation.
Reciprocally, the same assertion is applied for the understanding of the factorized covariant behavior of the helicity (spin-projection) at such quantum states.

\end{document}